\documentclass[prd,nofootinbib]{revtex4}
\usepackage{amsmath,amscd,amssymb}
\usepackage{epsfig}
\newcommand{\fract}[2]{{\textstyle\frac{#1}{#2}}}

\begin{document}

\title{Vacuum Polarization Energy for General Backgrounds
in One Space Dimension}

\author{H. Weigel}

\affiliation{
Physics Department, Stellenbosch University,
Matieland 7602, South Africa}

\begin{abstract}
For field theories in one time and one space dimensions we propose an 
efficient method to compute the vacuum polarization energy of static 
field configurations that do not allow a decomposition into symmetric and 
anti--symmetric channels. The method also applies to scenarios in which the 
masses of the quantum fluctuations at positive and negative spatial infinity 
are different. As an example we compute the vacuum polarization energy
of the kink soliton in the $\phi^6$ model. We link the dependence of this 
energy on the position of the soliton to the different masses.
\end{abstract}

\maketitle

\section{Introduction}

Vacuum polarization energies (VPE) sum the shifts of zero point energies of 
quantum fluctuations that interact with a (classical) background potential. 
Spectral methods~\cite{Graham:2009zz} have been very successful in computing VPEs 
particularly for background configurations with sufficient symmetry to facilitate 
a partial wave decomposition for the quantum fluctuations. In this approach
scattering data parameterize Green functions from which the VPE is determined. In 
particular the imaginary part of the two--point Green function at coincident 
points, {\it i.e.} the density of states, is related to the phase shift of 
potential scattering~\cite{Graham:2002xq}. Among other features, the 
success of the spectral methods draws from the direct implementation of background 
independent renormalization conditions by identifying the Born series for the 
scattering data with the expansion of the VPE in the strength of the potential.
The ultra--violet divergences are contained in the latter and can be re--expressed 
as regularized Feynman diagrams. In renormalizable theories the divergences are 
balanced by counterterms whose coefficients are fully determined in the 
perturbative sector of the quantum theory in which the potential is zero.

For field theories in one space dimension the partial wave decomposition 
separates channels that are even or odd under spatial reflection. We propose a 
very efficient method, that in fact is based on the spectral methods, to numerically 
compute the VPE for configurations that evade a decomposition into parity even 
and odd channels. This is particularly interesting for field theories that contain 
classical soliton solutions connecting vacua in which the masses of the quantum 
fluctuations differ. A prime example is the $\phi^6$ model. For this model some 
analytical results, in particular the scattering data for the quantum fluctuations, 
have been discussed a while ago in Refs.~\cite{Lohe:1979mh,Lohe:1980js}. 
However, a full calculation of the VPE has not yet been reported. A different 
approach, based on the heat kernel expansion with $\zeta$--function 
regularization~\cite{AlonsoIzquierdo:2011dy,AlonsoIzquierdo:2012tw} has already 
been applied to this model~\cite{AlonsoIzquierdo:2002eb}\footnote{See 
Refs.~\cite{Elizalde:1996zk,Elizalde:1994gf,Kirsten:2000ad} for reviews of heat kernel 
and $\zeta$--function methods.}. This approach requires an intricate formalism 
on top of which approximations (truncation of the expansion) are required. We 
will see that they become less accurate as the background becomes sharper. We 
also note that a similar problem involving distinct vacua occurs in scalar 
electrodynamics when computing the quantum tension of domain 
walls~\cite{Parnachev:2000fz}.

We briefly review the setting of the one--dimensional problem.
The dynamics of the field $\phi=\phi(t,x)$ is governed by the Lagrangian
\begin{equation}
\mathcal{L}=\frac{1}{2}\partial_\mu\phi\,\partial^\mu\phi-U(\phi)\,.
\label{eq:lag1}
\end{equation}
The self--interaction potential $U(\phi)$ typically has distinct minima and there may
exist several static soliton solutions that interlink between two 
such minima as $x\to\pm\infty$. We pick a specific soliton, say 
$\phi_0(x)$ and consider small fluctuations about it
\begin{equation}
\phi(t,x)=\phi_0(x)+\eta(t,x)\,.
\label{eq:fluc1}
\end{equation}
Up to linear order, the field equation turns into a Klein--Gordon type equation
\begin{equation}
\left[\partial_\mu\partial^\mu+V(x)\right]\eta(t,x)=0\,,
\label{eq:fluc2}
\end{equation}
where $V(x)=U^{\prime\prime}(\phi_0(x))$ is the background potential
generated by the soliton. At spatial infinity $V(x)$
approaches a constant to be identified as the mass (squared)
of the quantum fluctuations. In general, as {\it e.g.} for the 
$\phi^6$ model with
$U(\phi)=\frac{\lambda}{2}\phi^2(\phi^2-\Lambda^2)^2$, we allow
$\lim_{x\to-\infty}V(x)\ne\lim_{x\to\infty}V(x)$. This gives rise to 
different types of quantum fluctuations.
While $\phi_0$ is classical, the fluctuations are subject to canonical 
quantization so that the above harmonic approximation yields the leading 
quantum correction. As a consequence of the interaction with the background
the zero point energies of all modes change and the (renormalized) sum of 
all these changes is the VPE, {\it cf.} Sec. \ref{sec:VPE}.

\section{Phase Shifts}
\label{sec:phaseshifts}

As will be discussed in Sec. \ref{sec:VPE} the sum of the scattering 
(eigen)phase shifts is essential to compute the VPE from spectral methods. 
We extract scattering data from the stationary wave equation, 
$\eta(t,x)\to{\rm e}^{-iEt}\eta(x)$, 
\begin{equation}
E^2\eta(x)=\left[-\partial_x^2 +V(x)\right]\eta(x)\,.
\label{eq:stationary}
\end{equation}
According to the above described scenario we define $m_L^2=\lim_{x\to-\infty}V(x)$ 
and $m_R^2=\lim_{x\to\infty}V(x)$ and take the convention $m_L\le m_R$, otherwise 
we just relabel $x\to-x$. We introduce a discontinuous pseudo potential
\begin{equation}
V_p(x)=V(x)-m_L^2+\left(m_L^2-m_R^2\right)\Theta(x_m)
\label{eq:Vback}
\end{equation}
with $\Theta(x)$ being the step function. Any finite value may be chosen 
for the matching point $x_m$. In contrast to $V(x)$, $V_p(x)\to0$ as 
$x\to\pm\infty$. Then the stationary wave equation,~(\ref{eq:stationary}) 
reads
\begin{equation}
\left[-\partial_x^2 +V_p(x)\right]\eta(x)
=\begin{cases}k^2\eta(x)\,,\qquad & {\rm for}\quad x\le x_m\cr
q^2\eta(x)\,,\qquad & {\rm for}\quad x\ge x_m
\end{cases}
\label{eq:splitDEQ}
\end{equation}
where $k=\sqrt{E^2-m_L^2}$ and $q=\sqrt{E^2-m_R^2}=\sqrt{k^2+m_L^2-m_R^2}$.
We emphasize that solving Eq.~(\ref{eq:splitDEQ}) is equivalent to solving 
Eq.~(\ref{eq:stationary}). We factorize coefficient functions $A(x)$ 
and $B(x)$ appropriate for the scattering problem via
$\eta(x)=A(x){\rm e}^{ikx}$ for $x\le x_m$ and $\eta(x)=B(x){\rm e}^{iqx}$
for $x\ge x_m$: 
\begin{equation}
A^{\prime\prime}(x)=-2ikA^\prime(x)+V_p(x)A(x)
\qquad {\rm and}\qquad
B^{\prime\prime}(x)=-2iqB^\prime(x)+V_p(x)B(x)\,,
\label{eq:DEQAB}
\end{equation}
where a prime denotes a derivative with respect to $x$. In appendix~B
of Ref.~\cite{Graham:2002xq} related functions, $g_{\pm}(x)$ were introduced
to parameterize the Jost solutions for imaginary momenta. The boundary conditions
$A(-\infty)=B(\infty)=1$ and $A^\prime(-\infty)=B^\prime(\infty)=0$ yield the 
scattering matrix by matching the solutions at $x=x_m$. Above threshold, 
$k\ge\sqrt{m_R^2-m_L^2}$ so that $q$ is real, the scattering matrix is
\begin{equation}
S(k)=\begin{pmatrix}
{\rm e}^{-iqx_m} & 0 \cr 
0 & {\rm e}^{ikx_m}
\end{pmatrix}
\begin{pmatrix}
B & -A^\ast \cr
iqB+B^\prime & ikA^\ast-A^{\prime\ast}
\end{pmatrix}^{-1}
\begin{pmatrix}
A & -B^\ast \cr
ikA+A^\prime & iqB^\ast-B^{\prime\ast}
\end{pmatrix}
\begin{pmatrix}
{\rm e}^{ikx_m} & 0 \cr 
0 & {\rm e}^{-iqx_m}
\end{pmatrix}\,,
\label{eq:Smatrix}
\end{equation}
where $A=A(x_m)$, etc. are the coefficient functions at the matching point. 
Conventions are that the diagonal and off--diagonal elements of $S$ contain 
the transmission and reflections coefficients, respectively~\cite{Kiers:1996jt}.
Below threshold we parameterize for $x\ge x_m$: 
$\eta(x)=B(x){\rm e}^{-\kappa x}$ with $\kappa=\sqrt{m_R^2-m_L^2-k^2}\ge0$
replacing $-iq$ in Eq.~(\ref{eq:DEQAB}) so that $B(x)$ is real. Then
\begin{equation}
S(k)=-\,
\frac{A\left(B^\prime/B-\kappa-ik\right)-A^\prime}
{A^\ast\left(B^\prime/B-\kappa+ik\right)-A^{\prime\ast}}\,
{\rm e}^{2ikx_m}
\label{eq:Rcoef}
\end{equation}
is the reflection coefficient. In both cases we compute the 
sum of the eigenphase shifts $\delta(k)=-(i/2){\rm ln}{\rm det}S(k)$.
The negative sign on the right hand side of Eq.~(\ref{eq:Rcoef}) suggests 
that (in most cases) $\delta(0)$ is an odd multiple of $\frac{\pi}{2}$ in 
agreement with Levinson's theorem.  When the scattering problem 
diagonalizes into symmetric ($S$) and anti--symmetric ($A$) channels and
taking $\delta(k)\to0$ as $k\to\infty$, the 
theorem states that $\delta_S(0)=\pi(n_S-\fract{1}{2})$ and 
$\delta_A(0)=\pi n_A$, where $n_S$ and $n_A$ count the bound states in 
the two channels~\cite{Barton:1984py,Graham:2001iv}. The additional $-\pi/2$ 
in the symmetric channel arises because in that channel it is the derivative of 
the wave function that vanishes at $x=0$, rather than the wave function 
itself. For scattering off a background that does not decompose into these 
channels we have $\delta(0)=\pi(n-\fract{1}{2})$, where $n$ is the total 
number of bound states~\cite{Kiers:1996jt}. There are particular
cases in which $\delta(0)$ is indeed an integer multiple of $\pi$.
Examples are reflectionless potentials and the case $V(x)\equiv0$.
Then there exist threshold states contributing $\fract{1}{2}$ to $n$.

The step potential of hight $m_R^2-m_L^2$ centered at $x=x_m$ corresponds to 
$V_p\equiv0$. In this case the wave equation is solved by $A(x)=B(x)\equiv1$ and 
\begin{equation}
\delta_{\rm step}(k)=
\begin{cases}
(k-q)x_m\,,\qquad & {\rm for}\quad k\ge\sqrt{m_R^2-m_L^2}\cr 
kx_m-{\arctan}\left(\frac{\sqrt{m_R^2-m_L^2-k^2}}{k}\right)\,,
\qquad & {\rm for}\quad k\le\sqrt{m_R^2-m_L^2}
\end{cases}
\label{eq:deltaBar}
\end{equation}
agrees with textbook results.

\section{Vacuum Polarization Energy} 
\label{sec:VPE}

Formally the VPE is the sum of the shifts of the zero point 
energies due to the interaction with a background potential that is 
generated by the field configuration $\phi_0$,
\begin{equation}
E_{\rm vac}[\phi_0]=\frac{1}{2}\sum_j
\left(E_j[\phi_0]-E_j^{(0)}\right)+E_{\rm ct}[\phi_0]\,.
\label{eq:zero}
\end{equation}
Regularization for this logarithmically divergent sum is understood. When combined 
with the counterterms, $E_{\rm ct}$ a unique finite result arises after removing 
regularization. Typically there are two contributions in the sum of 
Eq.~(\ref{eq:zero}): (i) explicit bound and (ii) continuous scattering states. The 
latter part is obtained as an integral over one particle energies weighted by the
change in the density of states, $\Delta \rho(k)$. We find the density 
$\rho(k)=\frac{d N(k)}{dk}$ for scattering modes incident from negative 
infinity by discretizing $kL+\delta(k)=N(k)\pi$ where $\delta(k)$ is phase shift. 
Adopting the continuum limit $L\to\infty$ and subtracting the result from 
the non--interacting case yields the Krein formula~\cite{Faulkner:1977aa},
\begin{equation}
\Delta \rho(k)= \rho(k)-\rho^{(0)}(k)=\frac{1}{\pi}\frac{d}{dk}\delta(k)\,.
\label{eq:Krein}
\end{equation}
The situation for modes incident from positive infinity is not as straightforward. 
Here we count levels (above threshold) by setting $qL+\delta(k)=N(k)\pi$.
Since $k$ is the label for the free states we get an additional contribution 
to the change in the density of states
\begin{equation}
\frac{L}{\pi}\frac{d}{dk}\left[q-k\right]=
\frac{L}{\pi}\left[\frac{k}{\sqrt{k^2+m_L^2-m_R^2}}-1\right]=
\frac{L}{\pi}\left[\frac{\sqrt{E^2-m_L^2}}{\sqrt{E^2-m_R^2}}-1\right]\,.
\label{eq:omit}
\end{equation}
Formally it adds a portion to the VPE that is not sensitive to the 
details of the potential. Its omission corresponds to the selection of a 
particular $L$ independent part from the effective potential as {\it e.g.} 
in Eq.~(3.42) of Ref.~\cite{Parnachev:2000fz}.

Then the VPE is solely extracted from the Krein formula. Integrating by parts 
and imposing the no--tadpole renormalization prescription yields
\begin{equation}
E_{\rm vac} =
\frac{1}{2}\sum_j (E_j - m_L) - \frac{1}{2\pi} \int_0^\infty dk\,
\frac{k}{\sqrt{k^2+m_L^2}} \left(\delta(k)-\delta^{(1)}(k)\right) \,.
\label{eq:EvacReal}
\end{equation}
The explicit sum runs over the discrete bound states that are obtained 
from the solutions to Eq.~(\ref{eq:stationary}) that exponentially 
approach zero at spatial infinity. The subtraction under the 
integral refers to the Born approximation with respect to
the potential $V(x)-m_L^2$. We stress that it does not refer to
$V_p(x)$ because the no--tadpole renormalization implements a 
counterterm that is local in the full potential. In general this 
disallows to write $\delta^{(1)}(k)\sim-(1/2k)\int dx [V(x)-m_L^2]$,
because the Born approximation to the step potential cannot be 
written as this integral. Yet, its phase shift is well defined, 
Eq.~(\ref{eq:deltaBar}) and the large momentum contribution, which 
is represented by the Born approximation, can easily be computed 
from Eq.~(\ref{eq:deltaBar})
\begin{equation}
\delta_{\rm step}(k)\,\longrightarrow\, \frac{x_m}{2k}\left(m_R^2-m_L^2\right)
\qquad {\rm as}\qquad k\,\longrightarrow\,\infty\,.
\label{eq:deltaBarInfty}
\end{equation}
By definition, the Born approximation is linear in the  potential. We use 
Eq.~(\ref{eq:Vback}) to write 
$V(x)-m_L^2=V_p(x)+\left(m_R^2-m_L^2\right)\Theta(x_m)$ and obtain 
the Born approximation
\begin{equation}
\delta^{(1)}(k)=-\frac{1}{2k}\int_{-\infty}^\infty dx\, V_p(x)\Big|_{x_m}
+\frac{x_m}{2k}\left(m_R^2-m_L^2\right)
=-\frac{1}{2k}\int_{-\infty}^\infty dx\, V_p(x)\Big|_{0}\,.
\label{eq:Born}
\end{equation}
The subscript recalls that $V_p(x)$ is defined with respect to a specific 
matching point $x_m$. However, the final Born approximation does not depend 
on $x_m$. This is a step towards establishing that the VPE does not
depend on the matching point. We stress that this independence does
not reflect translational invariance of the system as described by shifting 
the coordinate $x\to x-x_0$ in $V(x)$. On the contrary, Eq.~(\ref{eq:Born}) 
shows that at least the Born approximation varies under this 
transformation\footnote{It seems suggestive that the Born approximation
should have a step function factor 
$\Theta(k-\mbox{\tiny $\sqrt{m_R^2-m_L^2}$})$. In the limit $m_R\to m_L$ its 
modification of the VPE is proportional to $\frac{x_m}{m_L}(m_R^2-m_L^2)^{3/2}$.
It is thus of higher order and also violates the $x_m$ independence. Hence
this factor is not part of the Born approximation.}.

When the potential is reflection symmetric the scattering problem separates 
into even and odd channels. This symmetry also implies $q=k$ and allows to 
analytically continue to imaginary $k=it$ with $t\ge0$  straightforwardly.
Integrating over $t$ collects the bound state contribution~\cite{Graham:2009zz}
and the VPE is 
\begin{equation}
E^{\rm (S)}_{\rm vac}=\int_{m_L}^\infty \frac{dt}{2\pi}\,
\frac{t}{\sqrt{t^2-m_L^2}}\,\left[
{\rm ln}\left\{g(t,0)\left(g(t,0)-\frac{1}{t}g^\prime(t,0)\right)\right\}
\right]_1\,.
\label{eq:EvacJost}
\end{equation}
Again the Born approximation has been subtracted as indicated by the
subscript. Here $g(t,x)$ is the non--trivial factor  of the Jost solution 
on the imaginary axis that solves the DEQ 
\begin{equation}
g^{\prime\prime}(t,x)=2tg^\prime(t,x)+V(x)g(t,x)
\label{eq:DEQJost}
\end{equation}
with the boundary condition $g(t,\infty)=1$ and $g^\prime(t,\infty)=0$.

Above we have used heuristic arguments to compute the VPE from scattering
data. We stress that it can be derived from fundamental concepts of
quantum field theory~\cite{Graham:2002xq}.

\section{Numerical Results}

For simplicity we scale to dimensionless coordinates and fields such that 
as many as possible model parameters, for example $\lambda$ and $\Lambda$ 
from the introduction, are unity.

In all considered cases we have ensured that the phase shift does 
not vary with the choice of $x_m$; that Levinson's theorem is
reproduced; and that attaching flux factors
$S_{11}\to \sqrt{\frac{q}{k}}\,{\rm e}^{i(q-k)x_m}\,S_{11}$
and $S_{22}\to \sqrt{\frac{k}{q}}\,{\rm e}^{i(k-q)x_m}\,S_{22}$
to the transmission coefficients always produces a unitary scattering 
matrix. When $m_L=m_R$ we have also numerically verified that 
the sum of the eigenphase shifts equals the phase of the transmission 
coefficient $S_{11}=S_{22}$~\cite{Aguirre:2016tin}.

\subsection{Symmetric background}

We first compare the result from the novel method for cases in 
which $V(x)$ is reflection symmetric and the approach 
via Eq.~(\ref{eq:EvacJost}) is applicable. Analytic results are 
available for the $\phi^4$ kink and sine--Gordon
models that have background potentials [as in Eq.(\ref{eq:fluc2})]
\begin{equation}
V_{\rm K}(x)=6{\rm tanh}^2(x)-2
\qquad {\rm and}\qquad
V_{\rm SG}(x)=8{\rm tanh}^2(2x)-4\,,
\label{eq:VKinkSG}
\end{equation}
with $m_L=m_R=2$. The numerical simulation for Eq.~(\ref{eq:EvacReal})
agrees with the respective VPEs, $E_{\rm vac,K}=\sqrt{2}/4-3/\pi$ and 
$E_{\rm vac,SG}=-2/\pi$~\cite{Ra82}, to better than one in a thousand.

We next compute the vacuum polarization energies of the
$U(\phi)=\fract{1}{2}(\phi^2+a^2)(\phi^2-1)^2$ model, where $a$ 
is a real parameter. For $a\ne0$ there is only a single 
soliton solution that interlinks the vacua\footnote{The potential
$U(\phi)$ has two global minima at $\phi=\pm1$ for $a^2>\fract{1}{2}$. 
When $a^2<\fract{1}{2}$ a third (local) minimum exists. The three minima 
are degenerate for $a=0$.} $\phi_{\rm vac}=\pm1$~\cite{Lohe:1979mh}:
\begin{equation}
\phi_0(x)=a\frac{X-1}
{\sqrt{4X+a^2\left(1+X\right)^2}}
\qquad {\rm where} \qquad X={\rm e}^{2\sqrt{1+a^2}\,x}\,.
\label{eq:sym6}
\end{equation}
For this model VPE results from a heat kernel 
calculation~\cite{AlonsoIzquierdo:2011dy} are available. By comparing to our 
results, we estimate the validity of the approximations
applied in the that approach. This comparison is essential because 
(to our knowledge) the only estimate of the VPE in the pure ($a=0$) 
$\phi^6$ model, which is a main target of the present investigation, utilizes 
this technique~\cite{AlonsoIzquierdo:2002eb}.
The results are presented in table~\ref{tab:symVPE} and we observe that the 
various computations agree well for moderate and large $a$. The methods
based on scattering data agree within numerical precision. But when $a$ is small
deviations of about 10-15\% are observed for the (approximative) heat kernel 
method.
\begin{table}
\centerline{
\begin{tabular}{c|c|c|c}
$a$ & heat kernel, Ref. \cite{AlonsoIzquierdo:2011dy} & 
Jost, Eq.~(\ref{eq:EvacJost}) & present, Eq.~(\ref{eq:EvacReal}) \cr
\hline
0.1 & -1.349 & -1.461 & -1.462 \cr
0.2 & -1.239 & -1.298 & -1.297 \cr
1.0 & -1.101 & -1.100 & -1.102 \cr
1.5 & -1.293 & -1.295 & -1.297
\end{tabular}}
\caption{\label{tab:symVPE}Numerical VPEs for the symmetric background 
based on the soliton of the $(\phi^2+a^2)(\phi^2-1)^2$ model.}
\end{table}

\subsection{Asymmetric background, identical vacua}

For the lack of a (simple) soliton model we consider the two parameter ($A$, $\sigma$)
pseudo potential $V_p(x)=Ax{\rm e}^{-x^2/\sigma^2}$. The present method can be applied 
directly but also the standard spectral methods, Eq.~(\ref{eq:EvacJost}) can employed 
after symmetrizing 
\begin{equation}
V_R(x)=A\left[(x+R){\rm e}^{-\frac{(x+R)^2}{\sigma^2}}
-(x-R){\rm e}^{-\frac{(x-R)^2}{\sigma^2}}\right]
\label{eq:symbg}
\end{equation}
so that the limit $R\to\infty$ should give twice the VPE 
of $V_p(x)$~\cite{Graham:1998kz}.
\begin{table}
\centerline{
\begin{tabular}{c|cccccc|c}
$R$ & 1.0 & 1.5 & 2.0 & 2.5 & 3.0 & 3.5 & present, Eq.~(\ref{eq:EvacReal}) \cr
\hline
$A=2.5\,,\,\sigma=1.0$ &
-0.0369 & -0.0324 & -0.0298 & -0.0294 & -0.0293 & -0.0292 & -0.0293 \cr
\hline\hline
$R$ & 4.0 & 5.0 & 6.0 & 7.0 & 8.0 & 9.0 & present, Eq.~(\ref{eq:EvacReal}) \cr
\hline
$A=0.2\,,\,\sigma=4.0$ &
-0.0208 & -0.0188 & -0.0170 & -0.0161 & -0.0158 & -0.0157 & -0.0157
\end{tabular}}
\caption{\label{tab:asymVPE}Comparison of different methods 
to compute the VPE for a non--symmetric background. The $R$ dependent data
are half the VPE of the background, Eq.~(\ref{eq:symbg}) computed 
via Eq.~(\ref{eq:EvacJost}).}
\end{table}
Table \ref{tab:asymVPE} verifies that agreement is obtained, but 
large values for $R$ are needed to avoid interference effects for 
wide background potentials.

\subsection{Asymmetric background, unequal vacua, $\phi^6$ model}

\begin{figure}
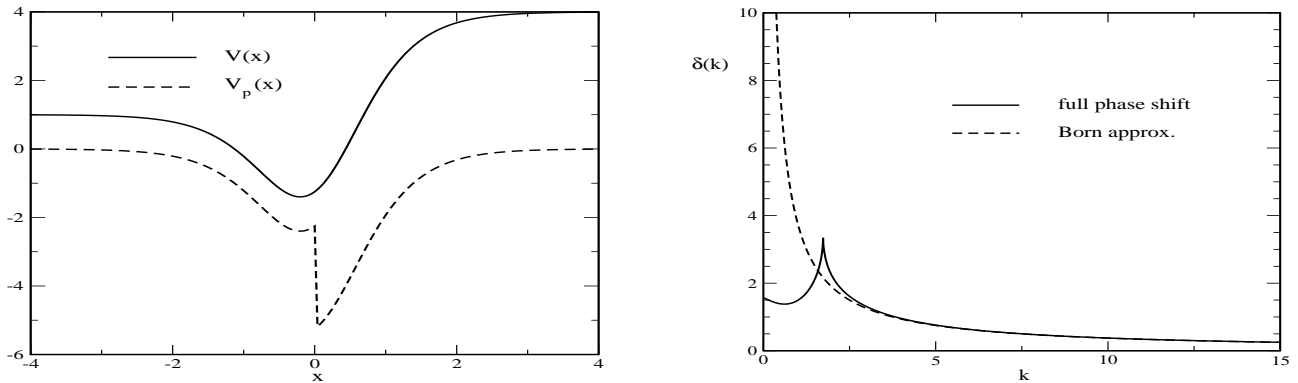

\centerline{
\includegraphics[width=8cm,height=5cm]{pot.eps}\hspace{1cm}
\includegraphics[width=8cm,height=5cm]{delta.eps}}
\caption{\label{fig:pot6}Potentials (left panel) and phase shift (right panel) for 
scattering off a soliton in the $\phi^6$ model. The pseudo potential $V_p(x)$ 
is shown for $x_m=0$.}
\end{figure}

We now turn to the pure $\phi^6$ model with 
$U(\phi)=\fract{1}{2}\phi^2\left(\phi^2-1\right)^2$. 
For $a=0$ the soliton of Eq.~(\ref{eq:sym6}) ceases to be a solution. 
However, there are solitons that interlink the degenerate vacua at 
$\phi_{\rm vac}=0$ and $\phi_{\rm vac}=\pm1$. The curvatures of $U(\phi)$ 
at these vacua differ so that the masses of the corresponding fluctuations 
are unequal. The soliton that corresponds to $m_L=1$ and $m_R=2$ is 
$\phi_0(x)=\left(1+{\rm e}^{-2x}\right)^{-1/2}$~\cite{Lohe:1979mh}. The resulting 
potentials for the fluctuations are shown in the left panel of figure \ref{fig:pot6}.
Also shown is the resulting sum, $\delta(k)$, of the eigenphase shifts as obtained from 
the scattering matrix, Eqs.~(\ref{eq:Smatrix}) and~(\ref{eq:Rcoef}). The direct numerical
calculation provides a discontinuous function between $-\pi/2$ and $\pi/2$. The 
discontinuities are removed uniquely by adding appropriate multiples of $\pi$ 
and demanding that $\delta(k)\to0$ as $k\to\infty$. In that limit it agrees with the
Born approximation, Eq.~(\ref{eq:Born}). However, the cusp, which is typical for 
threshold scattering, remains. Note also that $\delta(0)=\fract{\pi}{2}$ complies
with Levinson's theorem in one space dimension as there is only a single bound state: 
the translational zero mode of the soliton. 

Our results for the momentum dependence of the phase shift (and reflection coefficient) 
agree with the formulas given in Refs.~\cite{Lohe:1979mh,Lohe:1980js} up to overall signs. 
We are confident about our signs from Levinson's theorem and the Born approximation. Putting 
things together we find the vacuum polarization energy of the kink in the $\phi^6$ model
\begin{equation}
E_{\rm vac}=-0.5+0.4531=-0.0469\,,
\label{eq:final}
\end{equation}
where the summands denote the bound state and (renormalized) continuum 
parts as separated in Eq.~(\ref{eq:EvacReal}).

In Ref.~\cite{AlonsoIzquierdo:2002eb} the VPE of the $\phi^6$ model kink
was estimated relative to $V_\alpha(x)=\frac{3}{2}\left[1+{\rm tanh}(\alpha x)\right]$
for $\alpha=1$. In table \ref{tab:tanh} we give our results for various
values of $\alpha$. For $\alpha=1$ our relative 
VPE is $\Delta E_{\rm vac}=-0.0469-0.1660=-0.2129$ to be compared with
$-0.1264\sqrt{2}=-0.1788$ from Ref.~\cite{AlonsoIzquierdo:2002eb}. In 
view of the results shown in table \ref{tab:symVPE}, especially for small $a$, 
these data match within the validity of the approximations applied in the
heat kernel calculation.

\begin{table}
\parbox[b]{7.5cm}{
\begin{tabular}{c|ccccc|c}
$\alpha$ & 1.0 & 2.0 & 5.0 & 10.0 & 30.0 & step\cr
\hline
$E_{\rm vac}$& 0.1660 & 0.1478 & 0.1385 & 0.1363 & 0.1355 & 0.1355
\end{tabular}
\caption{\label{tab:tanh}
VPEs for $V_\alpha(x)=\fract{3}{2}[1+{\rm tanh}(\alpha x)]$.
The entry 'step' refers to using $\delta_{\rm step}$ from 
Eq.~(\ref{eq:deltaBar}) with $x_m=0$ in Eq.~(\ref{eq:EvacReal}).}
}\hspace{1cm}
\parbox[b]{8.5cm}{
\begin{tabular}{c|ccccc}
&\multicolumn{5}{c}{$E_{\rm vac}$}\cr
\hline
$x_0$& -2 & -1 & 0 & 1 & 2\cr
\hline
$\phi^6$ & 0.154 & 0.053 & -0.047 & -0.148 & -0.249\cr
$\alpha=2$ & 0.351 & 0.250 & 0.148 & 0.046 & -0.057\cr
$\alpha=5$ & 0.341 & 0.240 & 0.139 & 0.037 & -0.064
\end{tabular}
\caption{\label{tab:centerVPE}VPEs as a function of the center
of the configurations mentioned in the text. The two
entries $\alpha=2$ and $\alpha=5$ refer to the choices in
${\rm tanh}[\alpha(x+x_0)]$.}
}
\end{table}

\subsection{Translational variance and symmetrization}
\label{sec:trans}

We complete the discussion of the numerical results with a contemplation on 
translational invariance. In Sec.~\ref{sec:VPE} we have already seen that 
the Born approximation changes when the center of the configuration is 
shifted by a finite amount. To investigate this further, we compute the VPE 
for the $\phi^6$ kink $\left[1+{\rm e}^{-2(x+x_0)}\right]^{-1/2}$ in 
$U^{\prime\prime}(\phi)$ and $V(x)=\fract{3}{2}{\rm tanh}[\alpha(x+x_0)]$ 
as a generalization of the above study. The dependence on $x_0$ originates 
solely from the phase shift part because bound states move with $x_0$ without 
changing their energy eigenvalues. In Ref.~\cite{Lohe:1980js} this $x_0$ 
dependence was removed as part of the renormalization condition. This is not 
fully acceptable since the renormalization conditions should not depend on the 
field configuration.

For both potentials the numerical results from table \ref{tab:centerVPE} show 
that the VPE decreases by about 0.101 per unit of shifting the center towards 
negative infinity. We can build up a similar scenario in form of a symmetric 
barrier $V^{(x_0)}_{\rm SB}(x)=v_0\Theta\left(\frac{x_0}{2}-|x|\right)$ whose
VPE can be straightforwardly computed from Eq.~(\ref{eq:EvacJost}).
Substituting $V^{(x_0)}_{\rm SB}$ into the DEQ, 
Eq.~(\ref{eq:DEQJost}) yields
\begin{equation}
g(t,0)=\frac{\kappa_1{\rm e}^{-\kappa_2x_0/2}
-\kappa_2{\rm e}^{-\kappa_1x_0/2}}{\kappa_1-\kappa_2}
\qquad{\rm and}\qquad
g^\prime(t,0)=\frac{\kappa_1\kappa_2}{\kappa_1-\kappa_2}
\left({\rm e}^{-\kappa_2x_0/2}-{\rm e}^{-\kappa_1x_0/2}\right)\,,
\label{eq:barrSol}
\end{equation}
with $\kappa_{1,2}=t\pm\sqrt{t^2+v_0}$. Since we only consider the barrier 
with $v_0>0$, the $\kappa_{1,2}$ are always real. The relevant Born 
approximation is particularly simple
\begin{equation}
{\rm ln}\left\{g(t,0)\left(g(t,0)-\frac{1}{t}g^\prime(t,0)\right)\right\}
=\frac{v_0x_0}{2t}+\mathcal{O}\left(v_0^2\right)\,.
\label{eq:barrSolB1}
\end{equation}
We these ingredients we have evaluated the integral in Eq.~(\ref{eq:EvacJost})
using $v_0=m_R^2-m_L^2=3$ as suggested by the $\phi^6$ model kink and find 
\begin{equation}
\lim_{x_0\to\infty}\frac{E_{\rm vac}[V^{(x_0)}_{\rm SB}]}{x_0}\approx-0.1015\,.
\label{eq:barrRes}
\end{equation}
We can relate this result to the energy density of a step functuion potential
at spatial infinity using the phase shift from Eq.~(\ref{eq:deltaBar})
\begin{equation}
\frac{E_{\rm vac}[V^{(x_m)}_{\rm step}]}{|x_m|}\,\to\, -{\rm sign}(x_m)\,
\left[\int_0^{\sqrt{v_0}}\frac{dk}{4\pi}\,\frac{2k^2-v_0}{\sqrt{k^2+m_L^2}}
+\int_{\sqrt{v_0}}^\infty\frac{dk}{4\pi}\,
\frac{2k^2-2k\sqrt{k^2-v_0}-v_0}{\sqrt{k^2+m_L^2}}\right]
\qquad {\rm as}\quad |x_m|\to\infty\,.
\label{eq:Edensbarr}
\end{equation}
For $m_L=1$ and $v_0=3$ the expression is square brackets has the
numerical value $-0.1013$. These data suggest that
translational variance originates from the presence of the regions in 
which the quantum fluctuations have different masses. The numerical
results in table \ref{tab:centerVPE} and Eqs.~(\ref{eq:barrRes})
and~(\ref{eq:Edensbarr}) show that the rate at which the VPE changes 
is not sensitive to the particular shape of the background; but it 
depends on $v_0$.  Formally we could add the omission of Eq.~(\ref{eq:omit}) 
$$
\int \frac{dk}{2\pi}\, \sqrt{k^2+m_L^2}\,
\frac{d}{dk}\left[\sqrt{k^2-v_0}-k\right]
\sim \int \frac{dk}{2\pi}\,
\frac{k}{\sqrt{k^2+m_L^2}}\left[k-\sqrt{k^2-v_0}\right]
$$
to the energy density to eliminate the (leading) translational variance. The
above integration by parts misses a surface term whose divergence is regularized 
by the Born subtraction in the actual calculation of Eq.~(\ref{eq:Edensbarr}). 
We see that translational variance is qualitatively linked to the difference 
between the densities of states at positive and negative infinity, yet 
quantitative conclusions are not possible because that difference cannot be 
explicitly related to the center of the background potential. The picture
emerges that shifting the region with the larger mass towards negative
infinity removes modes from the spectrum and thus decreases the VPE. On 
the other hand it is not surprising that the bound state energies are 
translationally invariant because the bound state wave functions do not 
reach to spatial infinity.

By shifting the arguments in Eq.~(\ref{eq:VKinkSG}) we have verified that the
proposed numerical approach indeed produces translationally invariant VPEs
(actually phase shifts) for the $\phi^4$ and sine--Gordon solitons. In the
present formalism that verification is simple. In contrast, decoupling even and
odd parity channels, as required to obtain Eq.~(\ref{eq:EvacJost}), distinguishes
$x=0$ and does not leave space for varying the coordinate argument.

Substituting the symmetrized kink--antikink barrier 
\begin{equation}
\phi_0(x)=\left[1+{\rm e}^{2(x-\bar{x})}\right]^{-1/2}
+\left[1+{\rm e}^{-2(x+\bar{x})}\right]^{-1/2}-1
\label{eq:symkink}
\end{equation}
into $U^{\prime\prime}(\phi)$ produces a symmetric background that is a 
variation of a barrier with approximate width $2\bar{x}$. The vacuum is
characterized by $m_L=1$. Numerically we find
\begin{equation}
\lim_{\bar{x}\to\infty}
\left\{E_{\rm vac}[U^{\prime\prime}(\phi_0)]-2E_{\rm vac}[V^{(2\bar{x})}_{\rm SB}]\right\}
=-0.340=2\times (-0.170)
\label{eq:symkinkRes}
\end{equation}
which is in the right ball park in comparison with the data in the $\phi^6$ row of
table~\ref{tab:centerVPE}. Unfortunately, it is not clear which value of $x_s$ in 
$V^{(2x_s)}_{\rm SB}$ to use for the subtraction in Eq.~(\ref{eq:symkinkRes}). For 
example, it is sensible to define the center of the soliton 
$\left[1+{\rm e}^{2(x-\bar{x})}\right]^{-1/2}$ via its classical
energy density $\epsilon(x)=\fract{1}{2}\phi_0^{\prime2}+U(\phi_0)$:
$$
x_s=\frac{\int dx x \epsilon(x)}{\int dx \epsilon(x)}=\bar{x}+\fract{1}{2}
$$
and subtract $V^{(2x_s)}_{\rm SB}$ in Eq.~(\ref{eq:symkinkRes}). This changes 
that result to $-0.239=2\times (-0.120)$. When attempting to extract the kink 
VPE from the antikink--kink configuration
$\phi_0(x)=\left[1+{\rm e}^{-2(x-\bar{x})}\right]^{-1/2}
+\left[1+{\rm e}^{2(x+\bar{x})}\right]^{-1/2}$, a well of depth $v_0$ and width 
$2\bar{x}$ is generated yielding a completely different VPE due to the many bound 
states that emerge for large antikink--kink separation.

\section{Conclusion}

We have developed a method to compute the VPE for localized configurations
in one space dimension. It is based on spectral methods but generalizes 
previous approaches to configurations that are not amenable to a partial wave 
decomposition. Being a generalization of the spectral method, the novel
approach also naturally inherits the renormalization from the perturbative 
sector. The proposed method is very efficient: For a given background potential 
the numerical simulations only take only a few CPU minutes on a standard 
desktop computer. We solve two uncoupled second order ordinary differential 
equations, Eq.~(\ref{eq:DEQAB}), for the complex valued functions $A(x)$ and 
$B(x)$ that determine the scattering matrix. An equally simple 
equation~(\ref{eq:stationary}) yields the bound state energies.
Here we have only considered a 
single boson field, but taking $A(x)$ and $B(x)$ to be matrix valued 
straightforwardly  generalizes the method to multiple fields and/or fermions. 
The efficiency can also be established when confronting it with the heavy 
machinery needed for the heat kernel 
approach~\cite{AlonsoIzquierdo:2011dy,AlonsoIzquierdo:2002eb} that was 
earlier used to find the VPE of configurations lacking the symmetries for 
a partial wave decomposition. We consider the present method at least 
as efficient as that used in Ref.~\cite{Parnachev:2000fz}, which is based 
on a particular technique to compute functional determinants~\cite{Coleman}.
Both methods solve a differential equation for single particle energies. 
Integrating over these energies yields the VPE. 

As an application we have considered configurations for which the quantum 
fluctuations have different masses at positive and negative spatial infinity. 
Then the background can be interpreted as a modification of a step function 
potential that interpolates between different vacua. Though the parameterization 
of the solutions to the stationary wave equation differs on the left and right 
half lines (joined at the matching point $x_m$) we stress that we always solve 
the wave equation for the full problem. We have ensured that all results for 
the VPE (actually for the eigenphase shifts) do not depend on $x_m$. We did 
not explicitly compute the VPE versus another configuration; but the step 
function potential featured essential when (i) identifying the Born 
approximation for renormalization and (ii) establishing independence from 
technical parameters like $x_m$.

Though we may freely choose $x_m$ for computing the scattering
matrix, translational invariance with respect to the center of the soliton is 
lost when the masses of the quantum fluctuations differ at positive and 
negative spatial infinity. This loss of translational invariance signals that the 
global vacuum structure is locally relevant. We have also collected numerical and 
formal evidence that this position dependence is (mainly) due to the differences 
of the densities of states for scattering modes incident from positive or negative 
infinity. Even if this was the only cause, the multiple by which the corresponding 
spatial energy density should be subtracted is not unique leaving a residual position 
dependence.

In the $\phi^6$ model the exact no--tadpole renormalization scheme is required.
Any additional, though finite, renormalization of the counterterm coefficient is 
not well defined as the multiplying spatial integral is infinite.
However, this is not too surprising as the model is not fully renormalizable.

We wish to extend the present approach in the framework of the interface 
formalism~\cite{Graham:2001dy} and use it to investigate domain wall dynamics. This will 
allow a comparison with the results of Ref.~\cite{Parnachev:2000fz}. Also other
soliton models in one space dimension can be investigated. For example, the
$\phi^8$ model~\cite{Gani:2015cda} has solitons within different topological 
sectors. Comparing their VPE will shed some light on the relevance of 
quantum corrections to the binding energies of solitons that represent 
nuclei~\cite{Feist:2012ps}.

\section*{Acknowledgments}
Helpful discussions with N. Graham and M. Quandt are gratefully acknowledged.
This work is supported in parts by the NRF under grant~77454.


\begin{thebibliography}{99}

\bibitem{Graham:2009zz}
  N.~Graham, M.~Quandt, H.~Weigel,
  Lect.\ Notes Phys.\  {\bf 777} (2009) 1.

\bibitem{Graham:2002xq}
  N.~Graham, R.~L.~Jaffe, V.~Khemani, M.~Quandt, M.~Scandurra, H.~Weigel,
  Nucl.\ Phys.\ B {\bf 645} (2002) 49.

\bibitem{Lohe:1979mh}
  M.~A.~Lohe,
  Phys.\ Rev.\ D {\bf 20} (1979) 3120.

\bibitem{Lohe:1980js}
  M.~A.~Lohe, D.~M.~O'Brien,
  Phys.\ Rev.\ D {\bf 23} (1981) 1771.

\bibitem{AlonsoIzquierdo:2011dy}
  A.~Alonso--Izquierdo, J.~Mateos Guilarte,
  Nucl.\ Phys.\ B {\bf 852} (2011) 696.

\bibitem{AlonsoIzquierdo:2012tw}
  A.~Alonso--Izquierdo, J.~Mateos Guilarte,
  Annals Phys.\  {\bf 327} (2012) 2251.
  
\bibitem{AlonsoIzquierdo:2002eb}
  A.~Alonso--Izquierdo, W.~Garcia Fuertes, M.~A.~Gonzalez Leon, J.~Mateos Guilarte,
  Nucl.\ Phys.\ B {\bf 635} (2002) 525

\bibitem{Elizalde:1996zk}
  E.~Elizalde,
  Lect.\ Notes Phys.\ Monogr.\  {\bf 35} (1995) 1.

\bibitem{Elizalde:1994gf} 
  E.~Elizalde, S.~D.~Odintsov, A.~Romeo, A.~A.~Bytsenko, S.~Zerbini,\\
  {\it Zeta regularization techniques with applications},
  (World Scientific, Singapore, 1994).

\bibitem{Kirsten:2000ad} 
  K.~Kirsten,
  AIP Conf.\ Proc.\  {\bf 484}, 106 (1999).

\bibitem{Parnachev:2000fz}
  A.~Parnachev, L.~G.~Yaffe,
  Phys.\ Rev.\ D {\bf 62} (2000) 105034.

\bibitem{Kiers:1996jt}
  K.~Kiers, W.~van Dijk,
  J.\ Math.\ Phys.\  {\bf 37} (1996) 6033.

\bibitem{Barton:1984py}
  G.~Barton,
  J.\ Phys.\ A {\bf 18} (1985) 479.

\bibitem{Graham:2001iv}
  N.~Graham, R.~L.~Jaffe, M.~Quandt, H.~Weigel,
  Annals Phys.\  {\bf 293} (2001) 240.

\bibitem{Faulkner:1977aa}
  J.~S.~Faulkner, J. Phys. C, {\bf 10} (1977) 4661

\bibitem{Aguirre:2016tin}
  A.~R.~Aguirre, G.~Flores--Hidalgo,
  arXiv:1609.07341 [hep-th].

\bibitem{Ra82}
  R.~Rajaraman, {\it Solitons and Instantons}, North Holland, 1982

\bibitem{Graham:1998kz}
  N.~Graham, R.~L.~Jaffe,
  Phys.\ Lett.\ B {\bf 435} (1998) 145.

\bibitem{Coleman}
  S.~Coleman, {\it Aspects of Symmetry}, Cambridge University Press, 1985.

\bibitem{Graham:2001dy}
  N.~Graham, R.~L.~Jaffe, M.~Quandt, H.~Weigel,
  Phys.\ Rev.\ Lett.\  {\bf 87} (2001) 131601.

\bibitem{Gani:2015cda}
  V.~A.~Gani, V.~Lensky, M.~A.~Lizunova,
  JHEP {\bf 1508} (2015) 147.

\bibitem{Feist:2012ps}
  D.~T.~J.~Feist, P.~H.~C.~Lau, N.~S.~Manton,
  Phys.\ Rev.\ D {\bf 87} (2013) 085034.
\end{thebibliography}
\end{document}